\documentclass[12pt]{article}
\usepackage{graphicx}
\begin{document}

\title{The indication for $^{40}$K geo-antineutrino flux with Borexino phase-III data}

\author{L.~B.~Bezrukov,  I.~S.~Karpikov,  V.~V.~Sinev \\
Institute for Nuclear Research, Moscow, 117312  Russia}


\maketitle
\begin{abstract}
Abstract $-$ We provide the indication of high flux of $^{40}$K geo-antineutrino and geo-neutrino ($^{40}$K-geo-($\bar{\nu} + \nu$)) with Borexino Phase III data. This result was obtained by introducing a new source of single events, namely $^{40}$K-geo-($\bar{\nu} + \nu$) scattering on electrons, in multivariate fit analysis of  Borexino Phase III data. Simultaneously we obtained the count rates of events from $^7$Be, $pep$ and CNO solar neutrinos. These count rates are consistent with the prediction of the Low metallicity Sun model SSM B16-AGSS09. MC pseudo-experiments showed that the case of High metallicity Sun and absence of $^{40}$K-geo-($\bar{\nu} + \nu$) can not imitate the result of multivariate fit analysis of  Borexino Phase III data with introducing $^{40}$K-geo-($\bar{\nu} + \nu$) events. We also provide arguments for the high abundance of potassium in the Earth.  

\vspace{5mm}

\end{abstract}

\section{Introduction}

The Borexino collaboration reported in \cite{borex22} an improved measurement of the Carbon-Nitrogen-Oxygen (CNO) solar neutrino interaction rate at Earth obtained with the complete Borexino Phase-III dataset. The measured rate, $R_{\rm CNO} = 6.7^{+2.0}_{-0.8}$ cpd/100t (counts/(day $\times$ 100 tonnes)).

The main idea of this article is to see what could happen if we introduce in analysis the new source of single events from $^{40}$K  geo-antineutrino and neutrino ($^{40}$K-geo-($\bar{\nu} + \nu$)) scattering on electrons. Such source was proposed in \cite{bezruk, sinev}  and such analysis were done in \cite{bezrukPhase2} by using Borexino phase II dataset from \cite{borexsol} and  in \cite{
 bezrukPhase3.1} by using Borexino phase III dataset.

We will use in this paper the Phase-III dataset which is different from Phase-II dataset \cite{borexsol}.
The Phase-III dataset was taken when the radiopurity and thermal stability of the
detector was maximal, i. e. between January 2017 and 
October 2021. The Phase-II dataset was taken from July 2016 until February 2020. The Phase-III statistics is higher than the Phase-II statistics on about 30\%.

We hope that the obtained result will attract the attention of the scientific community to an alternative interpretation of the Borexino data and encourage further research of solar neutrino and potassium geo-antineutrino fluxes.

\section{Energy spectrum of recoil electrons from $^{40}$K-geo-($\bar{\nu} + \nu$)}

$^{40}$K abundance in natural mixture of K isotopes is 0.0117\%. It decays with $T_{1/2} = 1.248\cdot 10^{9}$ y through two modes: to $^{40}$Ca with efficiency 89.28\% emitting $\bar{\nu_{e}}$ with maximal energy 1.311 keV and to $^{40}$Ar emitting one of two possible mono-energetic $\nu_{e}$-s with energies 43.5 keV (10.67\%) and 1.504 keV ($\sim$0.05\%). Decay scheme of $^{40}$K can be found in \cite{dbase}. Low energy neutrinos cannot be detected but high energy ones can be seen in large volume detector as a small addition to antineutrino's effect.

\begin{figure*}[t!]
\begin{center}
\includegraphics[width=100mm]{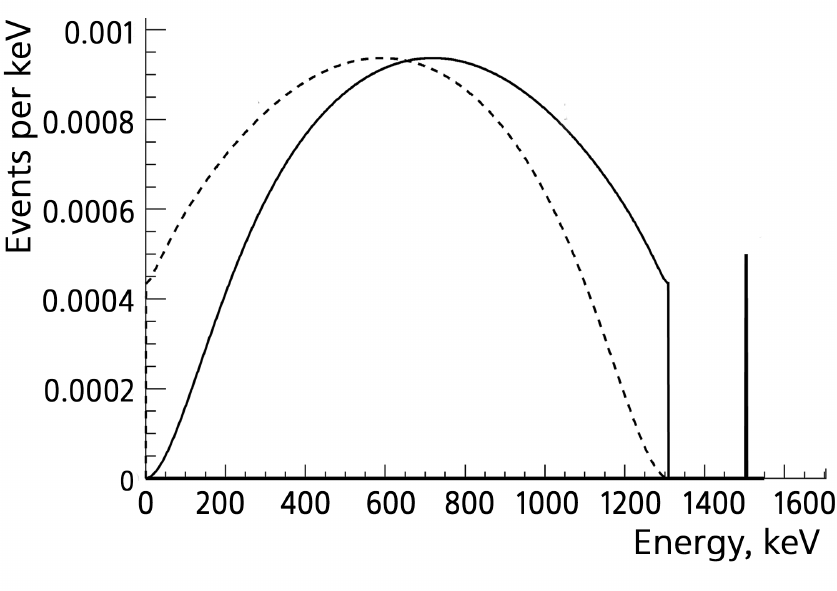}
\end{center}
\caption{\label{figzer} $^{40}$K antineutrinos and neutrinos spectra $-$ solid line. The spectra are normalized to the probablity of emission in decay: 0.8928 for $\bar{\nu_{e}}$-s and 0.0005 for $\nu_{e}$-s. Energy spectrum of electrons emitted in $^{40}$K decay $-$ dashed line.}
\end{figure*}

Figure \ref{figzer} demonstrates the total beta, antineutrino and neutrino energy spectra for $^{40}$K decay. The antineutrino spectrum was calculated using the beta one taken in \cite{dbase}. The beta spectrum is the same as was measured in \cite{kelly} and calculated in \cite{mougeot}.

\begin{figure*}[t!]
\begin{center}
\includegraphics[width=100mm]{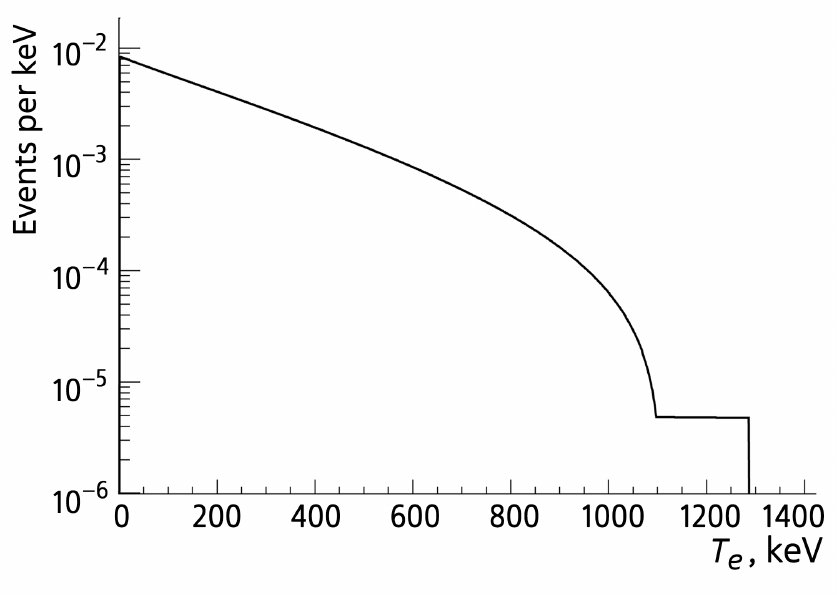}
\end{center}
\caption{\label{figone} The energy spectrum of recoil electrons from $^{40}$K-geo-($\bar{\nu} + \nu$). The pedestal at higher energies corresponds to neutrinos. The total counting rate of recoil electrons for this spectrum is 2.18 cpd/100t.}
\end{figure*}

In Figure \ref{figone} one can see the calculated energy spectrum of recoil electrons for $^{40}$K-geo-($\bar{\nu} + \nu$) that follows from the spectrum shape shown in Figure \ref{figzer}. The pedestal at higher energies corresponds to 1.5-MeV neutrinos. The total counting rate of recoil electrons for the spectrum shown in Figure \ref{figone} is equal to 2.18 cpd/100t that corresponds to potassium abundance 1.0\% of the Earth mass in the case of uniform distribution of potassium in the Earth.

The energy spectrum normalized to unit is the probability density function and we will use below the notation PDF for such normalized spectrum.

\begin{figure*}[t!]
\begin{center}
\includegraphics[width=100mm]{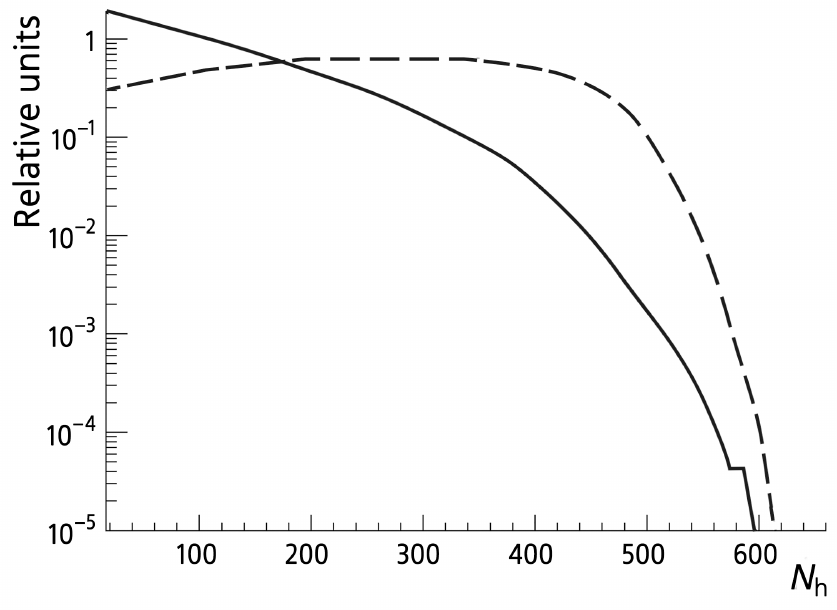}
\end{center}
\caption{\label{figtwo} Recoil electron spectrum from $^{40}$K-geo-($\bar{\nu}$ + $\nu$) transferred to $N_h$ (numder of hit PMTs)  scale according to the algorithm from \cite{borexphe}$-$ solid line. Energy spectrum of electrons emitted in $^{40}$K decay $-$ dashed line. }
\end{figure*}

We will use PDFs in multivariate fit analysis (MF) as a function of number of hit PMTs $N_h$ following the Borexino Collaboration. 
The calculated recoil electron energy spectrum from $^{40}$K-geo-($\bar{\nu} + \nu$) was transferred from the energy scale to  $N_h$ scale using the algorithm described in  \cite{borexphe}. In Figure \ref{figtwo} one can find $^{40}$K-geo-($\bar{\nu} + \nu$) PDF used in the MF analysis. $^{40}$K $\beta$ spectrum PDF is also shown here for comparison.

\begin{figure}[ht]
\begin{center}
\includegraphics[width=100mm]{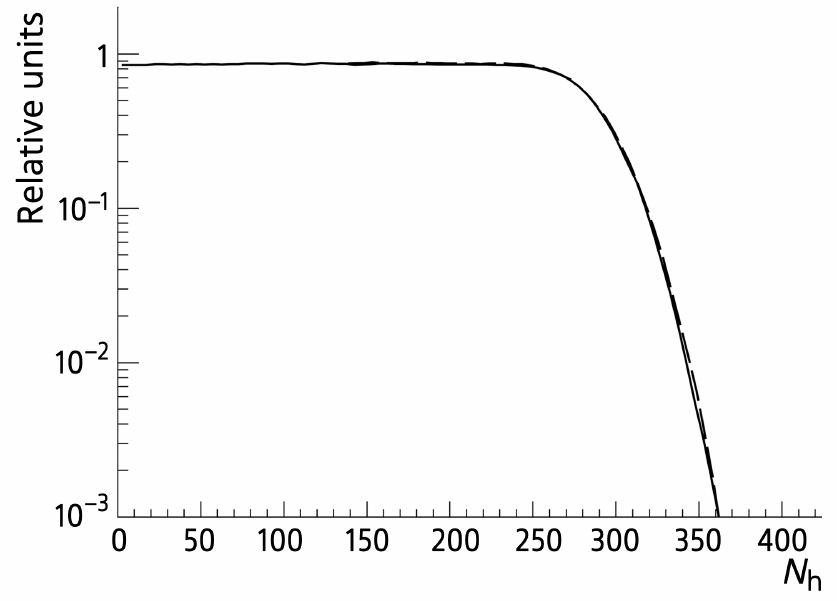}
\end{center}
\caption{\label{figbe}PDF of recoil electron spectrum from $^{7}$Be neutrinos transferred to $N_h$ (numder of hit PMTs) scale according to the algorithm from \cite{borexphe}. Solid line $-$ our calculation, dashed line $-$ \cite{borexsol}.}
\end{figure}

To prove that we correctly perform the calculation of $^{40}$K-geo-($\bar{\nu}$ + $\nu$) PDF, one can compare our $^{7}$Be PDF and the one taken from Borexino plot \cite{borexsol} shown in Figure \ref{figbe}.

\section{Energy spectrum of recoil electrons from CNO-$\nu$}

The CNO-$\nu$ spectrum consists of three components: the $^{13}$N, $^{15}$O and $^{17}$F neutrino spectra. The shape of these spectra is well known because the transitions are allowed and can be easily calculated. However, the fluxes depend on the solar model, and many flux predictions exist. In Table \ref{tabl:cnoflux} we can see a number of flux predictions for $^{13}$N, $^{15}$O and $^{17}$F according to different models.

\begin{table}[ht]
\caption{Neutrino fluxes produced in the Sun by nuclei $^{13}$N, $^{15}$O and $^{17}$F in units cm$^{-2}$ s$^{-1}$.}
\label{tabl:cnoflux}
\centering
\vspace{2mm}
\begin{tabular}{ c | c | c | c }
\hline
\hline
Model & $^{13}$N$\times10^{8}$ & $^{15}$O$\times10^{8}$ & $^{17}$F$\times10^{6}$ \\
\hline
B16  \cite{b16} & 5.03 & 1.34 & $<$8.5 \\
GS98  \cite{Villante} & 2.78 & 2.05 & 5.29 \\
AGSS09  \cite{Villante} & 2.04 & 1.44 & 3.26 \\
Caffau11  \cite{caffau11} & 2.801 & 2.123 & 4.648 \\
used in Borexino  \cite{vinoles} & 2.78 & 2.05 & 5.29 \\
\hline
\hline
\end{tabular}	
\end{table}

We have calculated neutrino spectra from $^{13}$N, $^{15}$O and $^{17}$F ourselves in the moment of production inside the Sun. These calculated spectra undergo oscillations according to the MSW-mechanism  \cite{micheev}. As a result each spectrum split in two: the electron neutrino spectrum and $\nu_{\mu}$ + $\nu_{\tau}$ spectrum. For each one we calculated a recoil electron spectrum and composed them in one. 
The sum of three CNO components appear similar to the CNO spectrum used in the Borexino analysis but not exactly. So, we decided to use in our analysis the CNO spectrum digitized from the plot of the Borexino publication. We can see  in Table \ref{tabl:cnoflux} that spectrum  used in the Borexino analysis corresponds to  High metallicity Sun model GS98.
 Later we are going to make the analysis using CNO spectrum corresponding to Low  metallicity Sun model AGSS09, see Table \ref{tabl:cnoflux}.

\section{Borexino experimental data  multivariate fit analysis}

We will use here the dataset digitized from the figure from \cite{borex22} with suppressed contribution of the cosmogenic $^{11}$C background. The histogram  of the number of events of Phase III $N$  depending on the number of hit PMTs  $N_h$  is shown in Table
\ref{tabl:bor3}.

\begin{table*}[ht]
\caption{Experimental histogram  of the number of events $N$ of Phase III depending on the number of hit PMTs $N_h$ from \cite {borex22}. Bin centers are shown in columns labeled as $N_h$. Events were collected over 871.31 days in 71.3 t.}
{\small
\label{tabl:bor3}
\vspace{2mm}
\begin{tabular}{|c|c|c| c | c | c | c | c | c | c | c | c | c | c | c | c | c | c | c | c |}
\hline
\hline
 $N_h$ & $N$ & $N_h$ & $N$ & $N_h$ & $N$ & $N_h$ & $N$ & $N_h$ & $N$ &  $N_h$ & $N$ & $N_h$ & $N$ & $N_h$ & $N$ & $N_h$ & $N$ \\
\hline
140&1021&230&931&320&167&410&55&500&54&590&34&680&25&770&23&860&40 \\
145&1098&235&834&325&123&415&68&505&46&595&34&685&22&775&35&865&36 \\
150&1251&240&725&330&110&420&43&510&53&600&31&690&22&780&26&870&37 \\
155&1573&245&663&335&105&425&56&515&50&605&31&695&20&785&34&875&39 \\
160&1801&250&677&340&76&430&71&520&50&610&26&700&15&790&31&880&35 \\
165&2097&255&602&345&712&435&62&525&39&615&25&705&20&795&33&885&28 \\
170&2614&260&797&350&71&440&43&530&44&620&31&710&20&800&34&890&35 \\
175&3044&265&555&355&60&445&46&535&47&625&21&715&14&805&37&895&32 \\
180&3341&270&562&360&55&450&65&540&36&630&21&720&19&810&42&900&26 \\
185&3471&275&479&365&59&455&57&545&36&635&30&725&24&815&26&905&33 \\
190&3341&280&425&370&56&460&45&550&26&640&28&730&24&820&36&910&25 \\
195&3285&285&406&375&51&465&62&555&35&645&25&735&24&825&38&915&22 \\
200&2869&290&365&380&44&470&71&560&46&650&16&740&19&830&38&920&27 \\
205&2443&295&316&385&59&475&70&565&35&655&25&745&24&835&29&925&22 \\
210&2151&300&265&390&49&480&58&570&39&660&17&750&29&840&39&930&20 \\
215&1741&305&233&395&46&485&52&575&35&665&25&755&22&845&33&935&13 \\
220&1368&310&191&400&51&490&40&580&24&670&25&760&24&850&44&940&19 \\
225&1160&315&198&405&56&495&57&585&37&675&23&765&32&855&52&945&11 \\
\hline
\hline
\end{tabular}
\hfill}
\end{table*}

We have calculated PDFs for most of the components used in the Borexino analysis and used them in our  multivariate fit (MF) analysis ($^{7}$Be, $pep$, $^{11}$C, $^{210}$Bi, $^{85}$Kr and CNO). Four components were digitized from  \cite{borexsol}, they are: $^{210}$Po alpha-peak, $^{8}$B and two backgrounds caused by external gammas from $^{208}$Tl and the summed spectrum of $^{214}$Bi and $^{40}$K gammas.

The $\chi^2$ function was used to estimate the goodness of our fit:

\begin{equation}
\chi^{2} = \sum_{i=1}^{162} \frac{(N_{i}-\sum_{k=1}^{10}w_{k}\cdot f_{k,i})^2}{\sigma^2_{i}},
\label{chi}
\end{equation}
where $N_{i}$ is the experimental value of counting rate of the Borexino detector events  in the $i$-th bin of  the number of hitted PMTs, $w_{k} = R_{k}\cdot 0.713\cdot t_{meas} -$ the weight for the PDF of the $k$-th component with $R_{k} -$ the total counting rate of the $k$-th component in cpd/100t, 0.713 $-$ the ratio of 71.3 t and 100 t and $t_{meas} -$ the measurement time in days, $f_{k,i} -$ the PDF of the $k$-th component and $\sigma_{i}$ is experimental uncertainty.
We used a standard tool for analysis, ROOT 6.22/08, to find the set of parameters $R_{k}\pm\Delta R_{k}$ which minimizes $\chi^{2}$. The $\Delta R_{k}$ corresponds to the confidence level of $68\%$ in ROOT 6.22/08. To check the results of ROOT we used also another minimization code.

\begin{table*}[ht]
\caption{Total counting rates and $\chi^{2}$ obtained without (column 1 and 1a) and with (column 2 and 2a) the $^{40}$K-geo-($\bar{\nu} + \nu$) component of the fit, cpd/100t. Comparison of the experimental results of solar neurtino counting rates from MFs and theoretical predictions for High metallicity (column 3) and Low metallicity (column 4)  \cite{borexphe}, cpd/100t.}
\small{
\label{tabl:bor4}
\centering
\vspace{2mm}
\begin{tabular}{|c|c|c|c|c|c|c|}
\hline
\hline
 &Mod.1,exp.&Mod.1,cpd range&Mod.2,exp.&Mod.2,cpd range&B16CS98&B16AGSS09\\
&&&&&HZ&LZ\\
\hline
&1&1 a&2&2 a&3&4\\
\hline
$^{7}$Be & 48.4 $\pm$ 0.9 &46$\div$49& 46.0 $\pm$ 0.84&42$\div$49 & 47.9 $\pm$ 2.8 & 43.7 $\pm$ 2.5 \\
$pep$ & 2.7 &2.7& 2.78 $\pm$ 0.07 &2.7$\div$2.78& 2.74 $\pm$ 0.04 & 2.78 $\pm$ 0.04 \\
CNO & 6.5 $\pm$ 0.7 &4.5$\div$8& 3.67 $\pm$ 0.73 &3.5$\div$20& 4.92 $\pm$ 0.55 & 3.52 $\pm$ 0.37 \\
$^{11}$C&1.7$\pm$ 0.1&1.35$\div$2& 1.83 $\pm$ 0.07 &1.4$\div$2&& \\
$^{210}$Po&41.5$\pm$0.4&40$\div$46&41.1$\pm$ 0.08&35$\div$45&& \\
$^{210}$Bi&11.8&9.8$\div$11.8& 11.0 $\pm$ 0.23&10.7$\div$11&& \\
$^{85}$Kr&8.5&3.5$\div$8.5&8.5 $\pm$ 0.24 &3.5$\div$8.5&& \\
$^{208}$Tl&4.76$\pm$ 0.2&4.3$\div$5& 4.76 &4.7/6&& \\
$^{214}$Bi&1.8$\pm$ 0.3&1.7$\div$2& 1.79 &1.79&& \\
$^{40}$Kgeo&0.0&0.0& 11 $\pm$ 0.6 &0$\div$11&& \\
\hline
 $\chi^{2}_1, \chi^{2}_2$&199.5&& 173.5 &&& \\
\hline
\hline
\end{tabular}
\hfill}
\end{table*}

\begin{figure*}[t!]
\begin{center}
\includegraphics[width=170mm]{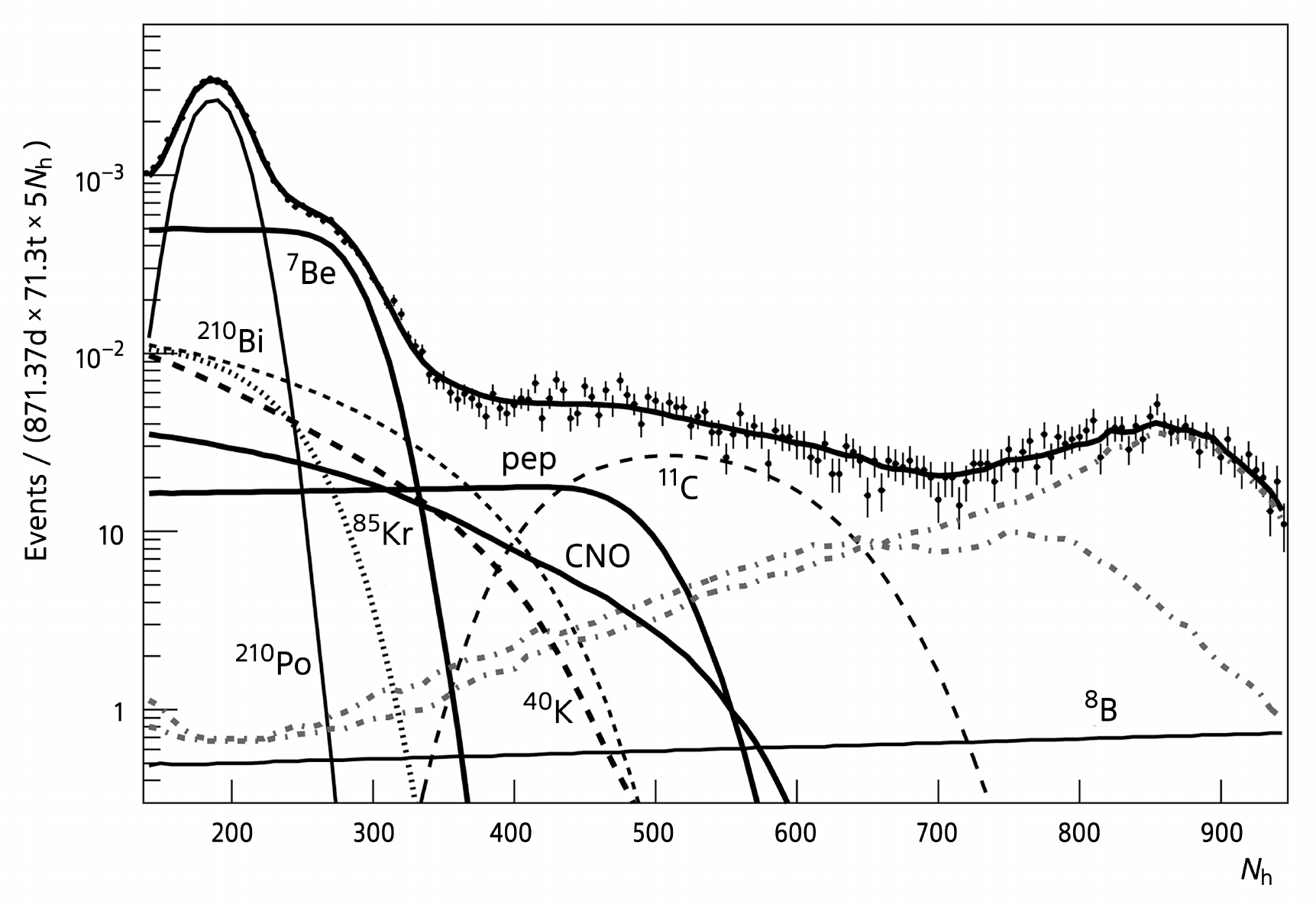}
\end{center}
\caption{\label{figfive} Counting rate of recoil electrons and alphas in Borexino detector in 5 hits bin versus number of hit PMTs. The values of points with errors from  \cite{borex22} are given in Table \ref{tabl:bor3}. The curves are the contributions of different components of the fit and the sum of all components obtained by the MF analysis. Dark black dashed curve is contribution of $^{40}$K-geo-($\bar{\nu} + \nu$) events. The obtained total counting rates for components of the fit are presented in the second column of the Table \ref{tabl:bor4}.}
\end{figure*}

The PDF for $^{40}$K-geo-($\bar{\nu}$ + $\nu$) was added to the analysis. So, in total we have 10 PDFs to fit: $^{7}$Be, $pep$, $^{11}$C, $^{210}$Po, $^{210}$Bi, $^{85}$Kr, external backgrounds 1 and 2, CNO and $^{40}$K-geo-($\bar{\nu}$ + $\nu$). $^{8}$B was fixed.

At the first stage we made the fit of Borexino dataset by minimization of $\chi^{2}$ from (\ref{chi}) using our calculated PDFs for the components but $^{40}$K-geo-($\bar{\nu}$ + $\nu$) was set to zero. We used the energy range from 140 to 945 hits following the Borexino analysis.
The experimental dataset was described with a sum of 8 variable components. The $pep$ was fix and other parameters were constrained as column 1a of Table \ref{tabl:bor4} shows. The upper limits of constraints for $^{210}$Bi and $^{85}$Kr we took from Borexino measurements of the upper limits of $^{210}$Bi and $^{85}$Kr concentrations in scintillator.
 As a result we obtained total counting rates that are on the first column of Table \ref{tabl:bor4}. We can observe the similarity of the obtained values of the parameters to the parameters presented by Borexino Collaboration in \cite{borex22, borexphe} and references inside them. The obtained values of $R$($^7$Be), $R(pep)$, $R$(CNO) support the Sun model of High metallicity. Theoretical predictions for High metallicity Sun are shown in column 3 of Table \ref{tabl:bor4}.

The PDF of $^{40}$K-geo-($\bar{\nu} + \nu$) was added at the second stage of the analysis. We made 8 parameters free (constrained) and external backgrounds 1 and 2 were fixed as column 2a of Table \ref{tabl:bor4} shows.

The result of this fit is shown in Figure \ref{figfive} and obtained total counting rates are in the second column of Table \ref{tabl:bor4}. We obtained the value of $R(^{40}$K-geo-$(\bar{\nu} + \nu)) = 11\pm 0.6$ cpd/100t. The fit found also the reasonable values of total counting rates for all components. The obtained values in second stage of ananlysis of $R$($^7$Be), $R(pep)$, $R$(CNO) support the Sun model of Low metallicity in contrary to the first stage analysis. Theoretical predictions for Low metallicity Sun are shown in column 4 of Table \ref{tabl:bor4}.
We note here that $\chi^{2}_2$ values obtained at the second stage is lower (better) than obtained at the first stage  $\chi^{2}_1$. The $\chi^{2}_1$ and $\chi^{2}_2$  values are shown in last row of Table \ref{tabl:bor4}.
This means that Low metallicity Sun and high abundance of potassiun in the Earth is more realistic than High metallicity Sun and low abundace of potassium in the Earth.

The inclusion of the scattering of $^{40}$K-geo-($\bar{\nu} + \nu$) on electrons in the analysis mainly reduces the number of events from $^{7}$Be and CNO. But the number of events from $pep$ becomes a little higher.

Note that the error for the counting rate of  $^{40}$K-geo-$(\bar{\nu} + \nu)$ events given in column 2 of  the Table \ref{tabl:bor4} characterizes the minimum of the function  $\chi^{2}$ and is not an experimental error in measuring the value of $R(^{40}$K-geo-$(\bar{\nu} + \nu))$.

\section{Simulation of the experiment}

We performed MC simulated experiments to understand the influence of the finite sample of data, the parameter constraints and the other conditions of the fit on reconstructed results. 
The MC simulated experiments help also to understand the possibilities of a next generation Borexino-type detector to measure the $^{40}$K-geo-$(\bar{\nu} + \nu)$ flux  
accurately.
 
We took a set of PDFs with certain fixed counting rates and simulated the Borexino-like spectrum of single events  many times via the MC method. Each time we had a new finite sample of Borexino-like data. We reconstructed the values of counting rates using the procedure (\ref{chi}) described above for each sample and, as a result, obtained probability distributions of the reconstructed counting rates and $\chi^{2}_1$ and $\chi^{2}_2$. We obtained wide non-symmetric distributions with their mean values different from the values fixed in MC mainly for components with low counting rates $R(^{40}$K-geo) and $R$(CNO). The main reason for this is the effect of statistical fluctuations of components with high counting rates.  These fluctuations affect the reconstructed counting rates of components with low counting rates. The value of this effect depends on statistics. Therefore, we found that the reconstructed counting rates in the Borexino-like data samples obtained using this procedure can have systematic biases relative to their true values.

Below we will give an examples of our pseudo-experiments.
We have two sets of components after our MFs of the experimental Borexino data sample. One set is in the first column of Table \ref{tabl:bor4} and another set is in the second column of Table \ref{tabl:bor4}. The main difference is the reconstructed $R(^{40}$K-geo-($\bar{\nu} + \nu)$), $R(^{7}$Be) and $R$(CNO) rates.

\begin{figure*}[t!]
\begin{center}
\includegraphics[width=120mm]{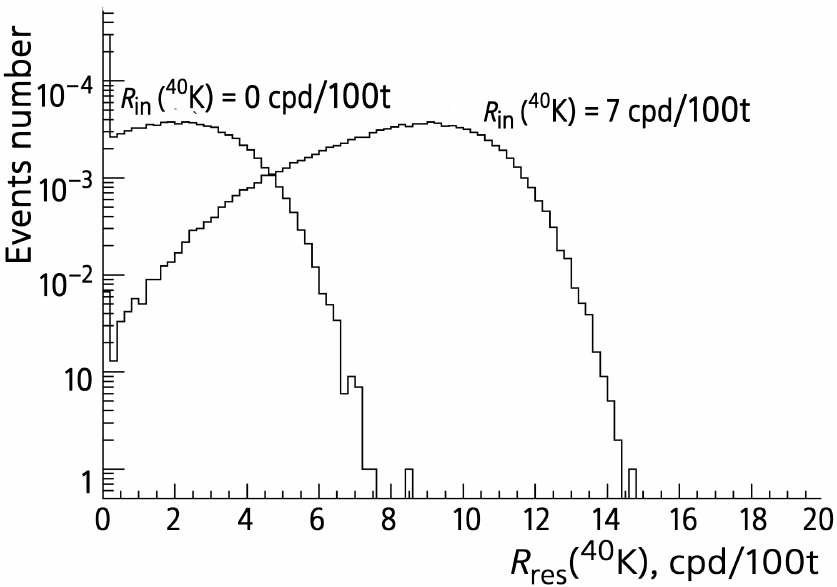}
\end{center}
\caption{\label{figmod} Reconstructed counting rate $R_{\rm{res}}(^{40}$K-geo-$(\bar{\nu} + \nu))$ distributions for $10^5$ pseudo-experiments. The curve labled $R_{\rm{in}}(^{40}$K) = 0 cpd/100t is the result of pseudo-experiments with parameters from column 1 of the Table \ref{tabl:bor4}. The mean value of this distribution is $R_{\rm{res}}(^{40}$K-geo-$(\bar{\nu} + \nu))$ = 1.67 cpd/100t and $\sigma = 1.53$ cpd/100t. The curve labled $R_{\rm{in}}(^{40}$K) = 7 cpd/100t is the result of pseudo-experiments with parameters from column 2 of the Table \ref{tabl:bor4} but with $R(^{40}$K-geo-$(\bar{\nu} + \nu))$ = 7 cpd/100t.}
\end{figure*} 

 At first we chose for our example of pseudo-experiments the parameters from first column of Table \ref{tabl:bor4}. But  when we reconstructed the parameters, we included the PDF of $R(^{40}$K-geo-($\bar{\nu} + \nu)$) in procedure (\ref{chi}) and applied the constraints as column 2a of Table \ref{tabl:bor4} shows. The idea was to test whether reality without potassium could simulate our result with $R(^{40}$K-geo-$(\bar{\nu} + \nu)) = 11$ cpd/100t.  The curve labled a $R_{\rm{in}}(^{40}$K) = 0 cpd/100t on Figure  \ref{figmod} shows the distribution of  reconstructed $R_{\rm{res}}(^{40}$K-geo-$(\bar{\nu} + \nu))$ for $10^5$ pseudo-experiments. We see that probability to observe $R(^{40}$K-geo-$(\bar{\nu} + \nu)) > 7$ cpd/100t  is less than $10^{-5}$.

 Then we chose for our another example of pseudo-experiments the parameters from second column of Table \ref{tabl:bor4}. Knowing about the existence of systematic biases, we took for MC simulation of data samples the counting rate $R(^{40}$K-geo-($\bar{\nu} + \nu)$) = 7 cpd/100t  instead of the value in the second column  of Table \ref{tabl:bor4}. We applied the constraints as column 2a of Table \ref{tabl:bor4} shows but  $R(^{40}$K-geo-$(\bar{\nu} + \nu))$ was free. 
 The curve labled $R_{\rm{in}}(^{40}$K) = 7 cpd/100t on Figure  \ref{figmod} shows the distribution of  reconstructed $R_{\rm{res}}(^{40}$K-geo-$(\bar{\nu} + \nu))$  for $10^5$ pseudo-experiments. We see that our result $R(^{40}$K-geo-$(\bar{\nu} + \nu)) = 11$  cpd/100t has the high probability in this case. We see also a systematic bias and the need a big statistic to measure the potassium abundance more accurately.  We  constructed the  $\chi^2_1$ and $\chi^2_2$ distributions obtained as a result of approximation these MC data samples by the sets of single event sources without of $R(^{40}$K-geo-$(\bar{\nu} + \nu))$ and with  of $R(^{40}$K-geo-$(\bar{\nu} + \nu))$ correspondently . The distributions has the gaussian form with $\bar{\chi}^2_1 = 184.4 \pm 21.17.$ and $\bar{\chi}^2_2 = 167.4 \pm 19.79.$ The obtained values $\chi^2_1 = 199.5$ and $\chi^2_2 = 173.5$ for experimental Borexino phase III dataset (see Table \ref{tabl:bor4}) is not contradict to these MC values.
  At first we chose for our example of pseudo-experiments the parameters from first column of Table \ref{tabl:bor4}. But  when we reconstructed the parameters, we included the PDF of $R(^{40}$K-geo-($\bar{\nu} + \nu)$) in procedure (\ref{chi}) and applied the constraints as column 2a of Table \ref{tabl:bor4} shows. The idea was to test whether reality without potassium could simulate our result with $R(^{40}$K-geo-$(\bar{\nu} + \nu)) = 11$ cpd/100t.  The curve labled a $R_{\rm{in}}(^{40}$K) = 0 cpd/100t on Figure  \ref{figmod} shows the distribution of  reconstructed $R_{\rm{res}}(^{40}$K-geo-$(\bar{\nu} + \nu))$ for $10^5$ pseudo-experiments. We see that probability to observe $R(^{40}$K-geo-$(\bar{\nu} + \nu)) > 7$ cpd/100t  is less than $10^{-5}$.

 Then we chose for our another example of pseudo-experiments the parameters from second column of Table \ref{tabl:bor4}. Knowing about the existence of systematic biases, we took for MC simulation of data samples the counting rate $R(^{40}$K-geo-($\bar{\nu} + \nu)$) = 7 cpd/100t  instead of the value in the second column  of Table \ref{tabl:bor4}. We applied the constraints as column 2a of Table \ref{tabl:bor4} shows but  $R(^{40}$K-geo-$(\bar{\nu} + \nu))$ was free. 
 The curve labled $R_{\rm{in}}(^{40}$K) = 7 cpd/100t on Figure  \ref{figmod} shows the distribution of  reconstructed $R_{\rm{res}}(^{40}$K-geo-$(\bar{\nu} + \nu))$  for $10^5$ pseudo-experiments. We see that our result $R(^{40}$K-geo-$(\bar{\nu} + \nu)) = 11$  cpd/100t has the high probability in this case. We see also a systematic bias and the need a big statistic to measure the potassium abundance more accurately.  We  constructed the  $\chi^2_1$ and $\chi^2_2$ distributions obtained as a result of approximation these MC data samples by the sets of single event sources without of $R(^{40}$K-geo-$(\bar{\nu} + \nu))$ and with  of $R(^{40}$K-geo-$(\bar{\nu} + \nu))$ correspondently . The distributions has the gaussian form with $\bar{\chi}^2_1 = 184.4 \pm 21.17.$ and $\bar{\chi}^2_2 = 167.4 \pm 19.79.$ The obtained values $\chi^2_1 = 199.5$ and $\chi^2_2 = 173.5$ for experimental Borexino phase III dataset (see Table \ref{tabl:bor4}) is not contradict to these MC values.

\begin{figure*}[t!]
\begin{center}
\includegraphics[width=120mm]{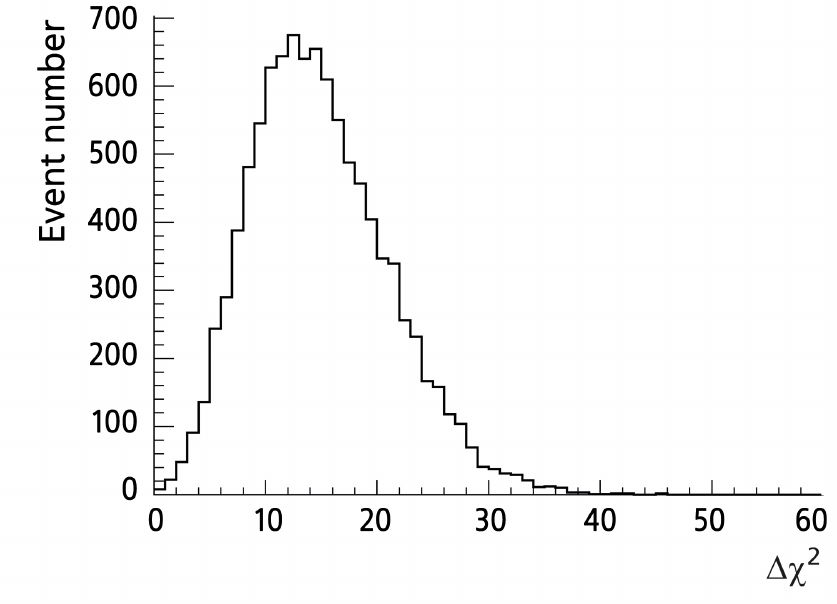}
\end{center}
\caption{\label{chi2} Distribution of $\Delta \chi^2 = \chi^2_1 - \chi^2_2$ for $10^4$ pseudo-experiments for which we chose the parameters from the second column of Table \ref{tabl:bor4} but  $R(^{40}$K-geo-$(\bar{\nu} + \nu))$ = 7 cpd/100t.
$\chi^2_1 -$  result of fit of pseudo-experiment dataset with the sum of event souces without  $^{40}$K-geo-$(\bar{\nu} + \nu)$.  $\chi^2_2 -$  result of fit of the same dataset but with  $^{40}$K-geo-$(\bar{\nu} + \nu)$.}
\end{figure*} 

     It is interesting also to compare the experimental $\Delta\chi^2 =  \chi^2_1 - \chi^2_2 = 26$ from Table \ref{tabl:bor4} with MC distribution of this value under assumption that Low metallicty Sun and the high  $^{40}$K-geo-$(\bar{\nu} + \nu)$ flux exists in nature. Figure  \ref{chi2} showes the MC distribution of $\Delta\chi^2$ for this case. The value $\Delta\chi^2  = 26$ is not contradict to this MC distributuon.

\section{Discussion}

The Phase III dataset is different from Phase II dataset. The main difference is in the energy dependence at low energies. This difference leads to different MF results in the case of inclusion in the analysis of  $^{40}$K-geo-$(\bar{\nu} + \nu)$ \cite{bezrukPhase2}. 

 We put the constraint for $R(^{40}$K-geo-$(\bar{\nu} + \nu)) \le 11$ cpd/100t (see column 2a of Table \ref{tabl:bor4}) and obtained  that $R(^{40}$K-geo-$(\bar{\nu} + \nu))$  is equal to upper limit of constraint and $R$(CNO) = 3.67 cpd/100t. This value is close to the prediction of LZ solar model. This is why we use here the constraint  $R(^{40}$K-geo-$(\bar{\nu} + \nu)) \le 11$ cpd/100t.  If we put $R(^{40}$K-geo-$(\bar{\nu} + \nu))$ free we can observe a further improvement of $\chi^2$, an increase in the reconstructed value $R(^{40}$K-geo-$(\bar{\nu} + \nu))$ and a decrease in $R$(CNO). 

Such behavior of the MF result for Phase III dataset  can be explained by the existence in the Earth of a huge amount of potassium.

 Taking into account the discovered bias of the reconstructed value, we propose for further discussion to adopt the following value of the counting rate:

\vspace{4mm}
\hspace{2cm} $R(^{40}$K-geo-$(\bar{\nu} + \nu)) = 7 ^{+2}_{ - 3}$ cpd/100t.   
\vspace{4mm}

This mean that abundunce of potassium in the Earth is in the range from 1.8\% to 4.1\% from the Earth mass.

\section{Radiogenic terrestrial heat production }

\hspace{0.5cm}

Suppose that the counting rate we have obtained $R(^{40}$K-geo-$(\bar{\nu} + \nu)) = 7$ cpd/100t is given by the actual $^{40}$K in the Earth. This means that about 3\% of the Earth's mass is potassium. 

Let calculate the intrinsic Earth heat flux for this potassium abundance.
The mass of $^{40}$K in the Earth in this case is:
$m(^{40}\rm{K}) = 2.1 \cdot 10^{22}$ g.

The equation relating masses and heat production (power)  is
\begin{equation}\label{eq:8}
H = m \cdot \frac{N_{Avog}}{A} \cdot \frac{E_{release}}{\tau} \cdot \alpha, 
\end{equation}
	where $N_{Avog}$ - Avogadro number, $A$ - atomic number, $E_{release}=0.6$ MeV - average energy release in $^{40}$K decay, $\tau = t_{1/2} / ln2$ - mean lifetime of isotope, $\alpha$ - the conversion factor $1\ \rm{MeV} = 1.6 \cdot 10^{-13}$ J
\begin{equation}\label{eq:9}
H(^{40}\rm{K}) = 2.1\cdot 10^{22}\rm{ g} \frac{6 \cdot 10^{23}}{40} \rm{g^{-1}} \cdot \frac{0.89 \cdot 0.6+0.1\cdot1.5} {1.8 \cdot 10^9 \cdot 3.15 \cdot 10^7 s} \cdot 1.6 \cdot 10^{-13} J = 616 \ TW. 
\end{equation}

Add the heat production from uranium and thorium following the work \cite{Ongeoneutrino}:

 $H(\rm{U}) = 16.1 \ TW$  and  $H(\rm{Th}) = 18 \ TW$.
\begin{equation}\label{eq:10}
H = H({\rm{U}}) + H({\rm{Th}}) + H(^{40}\rm{K}) = 650 \ \rm{TW}. 
\end{equation}

This estimated value is rather high, but it is consistent with predictions of Earth model "Hydridic Earth model (HE model)" \cite{Larin, toul}. HE model predicts the high potassium abundance (up to several persent of Earth mass).

The obtained value of heat production \ref{eq:10} is enough to explain the observed by ARGO project the increase of the ocean temperature \cite{Argo}. Enegy release in 300 TW during 30 years is enough to heat the ocean as ARGO observed.
 In the frame of HE model extra heat produced inside can be absorbed by the Earth itself through its expansion.
 Latter leads to the Earth cooling. The cause of the Earth expansion is the following. The radiogenic heat of the Earth interior leads to hydride decomposition. Protons and metal appear as a result. The metal volume is larger than volume of initial hydrides. This is the reason of Earth expansion.      

How much energy is needed to enhance  the Earth radius by $h = 1$ cm?  
\begin{equation}\label{eq:11}
M_m\cdot g\cdot h =  4.91\cdot10^{24}\rm{kg}\cdot 6\ \rm{\frac{m}{s^2}}\cdot 10^{-2}m = 3\cdot10^{23}J,
\end{equation}
here $M_m$ is the Earth mantle mass.

What work can be done by the energy release 350\ \rm{TW} during 30 years?
\begin{equation}\label{eq:12}
3.5\cdot10^{14}\rm W\cdot3.15 \cdot 10^{7} s \cdot30 y =  3.3\cdot 10^{23}J. 
\end{equation} 

One  can see that the expressions \ref{eq:11} and \ref{eq:12} are comparable. So, if the Earth radius would be increaesd by more than 1  cm, then the decomposition of hydrides be stopped and the cooling wave starts to spread through the  Earth's body up to surface.  This can cause a new ice age on the surface or make decreasing of mean temperature at least.  

\section{ Arguments in favor of the high potassium abundace in the Earth}

\vspace{0.5cm}
\hspace{0.5cm} The widespread belief in the fairness of Silicate Earth model and belief in the validity of the results of work \cite{davis} that the heat production in the Earth interior is equal to 47$\pm$2 TW do not allow to consider our result as the evidence of high flux of $^{40}$K-geo-($\bar{\nu} + \nu$).

Often the HE model is critisized by the following points.
 The entire Earth  have to be melted due to radiogenic heat and spent the most part of its life in this state if the Earth contains potassium more than 1\% of the Earth mass. Also the HE model predicts that the current heat production in the Earth's interior can be 200 TW and more which contradicts to the result of the work \cite{davis}.

However,
these arguments are not fully reliable. In particular, the entire Earth could not be melted because the HE model contains a subsurface cooling mechanism. This mechanism is activated when the subsurface is heated enough to decompose the metal hydrides. Therefore, in the HE model, the subsurface temperature oscillates \cite{Larin} and does not grow up till hydrides exist in the Earth. This argument was used in \cite{bezr2, bezr2018, barab2019, bezr22018}. It is noted in these works that thermal conductivity is not the main mechanism of heat transfer in the Earth, but hot protons and hydrogen-containing gases carry out the heat away. In these works the experimental evidences are provided that the heat production in Earth interior can reach the several hundreds TW. These are the heating of the oceans \cite{Argo}, the temperature profile of ultra-deep wells and non-direct evidence – the heat production in  Moon interior.

\section{Next importat stepts in the Earth studies}

Our analysis showed that the potential capabilities of the Borexino-type detector are not exhausted. 

Let consider a new detector of the Borexino-type with lower backgrounds and higher statistics. 
This require the low background nylon for detector inner vessel, better energy resolution, and detector must be plasced  deeper underground. 

As a result of pure nylon inner vessel, a larger fiducial volume could be achieved due to smaller amount of $^{210}$Po emanating into the scintillator. The latter will make it possible to measure the concentration of $^{210}$Bi in the scintillator by studying the Low Polonium Field in a stationary scintillator.
Exact  knowledge of bismuth content in the scintillator will reduce the uncertainty of the reconstructed values of the $^{40}$K-geo-($\bar{\nu} + \nu)$ counting rate.

\section{Conclusion}

We carried out our own analysis of the Borexino experimental data published in Ref. \cite{borex22}. 

\begin{itemize}
\item We calculated the recoil electron energy spectrum from $^{40}$K-geo-($\bar{\nu} + \nu$) scattering on electrons and transferred it to the numder of hit PMTs  scale.

\item We intodused a new source of single events, namely $^{40}$K-geo-($\bar{\nu} + \nu$) scattering on electrons, in multivariate fit analysis.

\item We provide the indication of high flux of $^{40}$K-geo-($\bar{\nu} + \nu$) with Borexino phase III data. The abundance of potassium  in the range $2\div4\%$ from the Earth mass can give such flux.

\item We obtained the count rates of $^7$Be, $pep$ and CNO solar neutrinos. These count rates are consistent with the prediction of the Low metallicity Sun model SSM B16-AGS09.

\item MC pseudo-experiments showed that the High metallicity Sun and absence of $^{40}$K-geo-($\bar{\nu} + \nu$) can not simulate  the result of  multivariate fit analysis with inrtodusing of $^{40}$K-geo-($\bar{\nu} + \nu$) events in analysis.

\item We consider our results as a support of Hydridic Earth model. We discussed the problem of high thermal radiogenic flux from the Earth interior in the case of the observed potassium abundance. 

\item We proposed to bild the next generation Borexino-type detector to measure $^{40}$K-geo-($\bar{\nu} + \nu$) flux with better accuracy .

\end{itemize}

\vspace{2mm}
\section*{Acknowledgments}

We are grateful to G. V. Sinev for valuable discussions on transferring energy loss to observable number of photoelectrons and idea to generate pseudo-experiments of the Borexino data, to F. L. Bezrukov for discussions and fruitful remarks and to I. I. Tkachev for common support and the possibility to discuss the results in his seminar.


\end{document}